\input{aipcheck}

\documentclass[
    ,final            
  ]
  {aipproc}

\usepackage{graphicx}

\layoutstyle{6x9}

\begin{document}

\title{Exact Stochastic Mean-Field dynamics}

\classification{03.65.Yz, 05.10.Gg, 05.70.Ln}
\keywords      {Open Quantum Systems, N-body problem, Stochastic methods}

\author{Denis Lacroix}{
address={GANIL, CEA and IN2P3, Bo\^ite Postale 5027, 14076 Caen Cedex, France},
email = {lacroix@ganil.fr}
 }

\iftrue
\author{Guillaume Hupin}{
address={GANIL, CEA and IN2P3, Bo\^ite Postale 5027, 14076 Caen Cedex, France} 
}

\begin{abstract} 
The exact evolution of a system coupled to a complex environment can be described by a 
stochastic mean-field evolution of the reduced system density. The formalism developed in 
Ref. \cite{Lac08} is illustrated in the Caldeira-Leggett model where a harmonic oscillator is coupled 
to a bath of harmonic oscillators. Similar exact reformulation could be used to extend mean-field transport theories 
in Many-body systems and incorporate two-body correlations beyond the mean-field one. The connection between open quantum 
system and closed many-body problem is discussed. 
\end{abstract}
\maketitle


\section{Introduction}

Dynamical Mean-field theories, such as Time-Dependent Hartree-Fock (TDHF), 
provide a suitable microscopic framework to treat both nuclear structure and reactions. Several limitations of mean-field theory exist so far. First, important correlations related to direct two-body effects are missing. Closely related to this problem, TDHF generally underestimates quantum fluctuations in collective degrees of freedom. During the past decades, several approaches have
been developed to incorporate correlations beyond mean-field. Stochastic methods, which uses the connection between open quantum system and the reduced description of a closed system in terms of few degrees of freedom, are among the most promising \cite{Abe96,Lac04}. Up to now, stochastic transport theories were proposed to treat approximately effects beyond the mean-field.  
Recently, we have shown that stochastic mean-field theories could provide an {\it exact} reformulation of both the quantum N-body problem 
\cite{Lac07} and/or open quantum systems\cite{Lac08}. In this proceedings, basic ingredients of the theory are exposed. The technique is then 
illustrated in the Caldeira-Leggett model. 
 
\section{Exact stochastic methods for Open Quantum systems}

We consider here a system (S) + environment (E) described by a Hamiltonian
\begin{eqnarray}
H = h_S + h_E + h_I, 
\end{eqnarray}   
where $h_S$ and $h_E$ denote the system and environment Hamiltonians respectively 
while $h_I$ is responsible for the coupling. Here we assume that the interaction Hamiltonian is written as 
$h_I =  \mathbf{Q} \otimes \mathbf{B}$. 

For the sake of simplicity, we assume 
an initial separable density $D(t_0) = \rho_S(t_0) \otimes \rho_B(t_0)$. As discussed in \cite{Lac08},
this assumption could eventually be relaxed. 
The exact S+E
evolution of the system is described by the Liouville von-Neumann equation $i\hbar \dot D = \left[H,D\right].$
Due to the coupling, the simple separable 
structure of the initial condition is not preserved in time. However, the exact density 
of the total system $D(t)$ could be obtained as an average over
simple separable densities, i.e. $D(t) = \overline{\rho_S(t) \otimes \rho_B(t)}$. {
In its simplest version, the stochastic process takes the form  \cite{Sha04,Lac05}
\begin{eqnarray}
\left\{
\begin{array} {lll}
d\rho_S &=& \frac{dt}{i\hbar}[h_S ,\rho_S] + du_S \{ {\mathbf Q} , \rho_S \} + dv_S [ {\mathbf Q}  , \rho_S ] 
\\
\\
d\rho_E &=& \frac{dt}{i\hbar}[h_E,\rho_E]+ du_E \{ {\mathbf B} , \rho_E \} + dv_E [ {\mathbf B}  , \rho_E ] 
\end{array}
\right.
\label{eq:stocmfsimple}
\end{eqnarray} 
where $\{ . , . \}$ denotes the anti-commutator while $du_{S/E}$ and $dv_{S/E}$ are Gaussian stochastic variables 
with zero mean value and variances equal to:
\begin{eqnarray}
\begin{array}{ccc}
\overline{du_Sdu_E} &=& \overline{dv_Sdv_E} = \frac{dt}{2\hbar},~~~~
\overline{du_Sdv_E} = \overline{dv_Sdu_E} = 0. 
\end{array}
\label{eq:noiseusue}
\end{eqnarray}
Direct numerical application of Eq. (\ref{eq:stocmfsimple}) is often difficult due to the following reason: (i) 
the evolution of the environment can rarely be fully followed in time due to its complexity; (ii) the number of trajectories required 
to accurately describe the system by these equations is very large. To avoid these difficulties, mean-field theory can be introduced 
prior to the noise term. the mean-field evolution of a system+environment can be seen as the best approximation of the dynamics assuming 
that $D$ remains separable. This amount to replace $h_S$ by $h_S + \left\langle \mathbf{B}(t) \right \rangle_E \mathbf{Q}$ in the deterministic 
part of the system evolution. The noise term has to be modified accordingly leading to the reduced system evolution:
\begin{eqnarray} 
d \rho_S = \frac{dt}{i\hbar} \left[ h_S + \left\langle \mathbf{B}(t) \right \rangle_E \mathbf{Q}, \rho_S \right] + du_S \{ \mathbf{Q} 
-\langle \mathbf{Q} (t) \rangle_S  , \rho_S \}_+ 
- i dv_S [\mathbf{Q} -\langle \mathbf{Q} (t) \rangle_S , \rho_S].   
\label{eq:rhos}
\end{eqnarray}
Introduction of mean-field reduces significantly the numerical effort. Last equation also shows that the influence of the environment
is entirely contained in 
$\left\langle \mathbf{B}(t) \right \rangle_E \equiv Tr(\mathbf{B} \rho_E)$. In practice, it will be easier to follow this quantity in time 
instead of the full environment density given by Eq. (\ref{eq:stocmfsimple}). It has been shown in ref. \cite{Lac08} that 
$\left\langle \mathbf{B}(t) \right\rangle_E $ could be expressed exactly as  
\begin{eqnarray}
\left\langle \mathbf{B}(t) \right\rangle_E &=& {\rm Tr}(\mathbf{B}^I(t -t_0) \rho_E(t_0)) \nonumber \\
&-&\frac{1}{\hbar}\int^t_0 D(t,s) \left\langle \mathbf{Q} (s) \right\rangle_S ds 
-\int^t_0 D(t,s) du_E(s)  + \int^t_0 D_1(t,s) dv_E(s) . 
\label{eq:btime}
\end{eqnarray}
where $\mathbf{B}^I(t-s) \equiv U^\dagger_E(t,s) \mathbf{B} U_E(t,s)$, where $U_E$ denotes the environment propagator, while $D$ and $D_1$ are correlations functions of the 
environment defined by:
\begin{eqnarray}
D(t,s) & \equiv & i \langle [ \mathbf{B}, \mathbf{B}^I(t-s)] \rangle_E \label{eq:d1},
~~~~ 
D_1(t,s) \equiv \langle  \{ \mathbf{B}  - \langle \mathbf{B}(s)\rangle_E , \mathbf{B}^I(t-s)  \} \rangle_E \label{eq:d2}.
\end{eqnarray}
The method described here is exact. As illustrated below, it could provide a practical solution to open quantum systems 
when "simple" expressions of the correlations functions exist. Several general remarks can be drawn from Eq. (\ref{eq:rhos}-\ref{eq:btime}).
First, the mean-field entering in the system evolution is highly non-local in time and contains a stochastic part. Second, averaging over 
the different trajectories leads to $i\hbar \partial_t \overline{\rho_S} = \overline{[ h_S + \left\langle \mathbf{B}(t) \right \rangle_E \mathbf{Q}, \rho_S ]}$. We therefore conclude that the dynamics of an open quantum system can always be written as an average 
over mean-field evolutions. This conclusion is by itself highly non-trivial.   
   
\section{Application to Caldeira-Leggett model}
As an illustration, we consider the Caldeira-Leggett model \cite{Cal83} where a single harmonic oscillator is coupled to an environment 
of harmonic oscillators initially at thermal equilibrium, i.e. 
\begin{eqnarray}
h_S = \frac{P^2}{2M} + \frac{1}{2}M\omega_0^2 Q^2, ~~~~ h_E = \sum_{n}\left(\frac{p^2_n }{2m_n} + \frac{1}{2}m_n \omega^2_n x^2_n\right)
\end{eqnarray}
and $\mathbf{B} \equiv -\sum_n \kappa_n x_n$ \cite{Bre02}. In that case, $D$ and $D_1$ 
identify with the standard correlation functions \cite{Lac08}: 
\begin{eqnarray}
\hspace*{-0.2 cm}
D(\tau) &=&  2\hbar \int^{+\infty}_0 d\omega J(\omega) \sin(\omega \tau) , \label{eq:d1exp} \\
\hspace*{-0.2 cm} D_1(\tau) &=&  2\hbar \int^{+\infty}_0 d\omega J(\omega) \coth({\hbar \omega / 2k_B T}) \cos(\omega \tau) , \label{eq:d2exp}
\end{eqnarray}
where $J(\omega)\equiv \sum_n \frac{ \kappa^2_{n}} {2m_n \omega_n} \delta(\omega - \omega_n)$  
denotes the spectral density \cite{Bre02}. 

To solve Eq. (\ref{eq:rhos}) we take advantage of the fact that an initial Gaussian density remain Gaussian in time due to the Harmonic
nature of $h_S$ and the specific coupling. Therefore, the density evolution can be replaced by its first and second moments evolution, given
by:   
\begin{eqnarray}
\left\{
\begin{array} {lll}
d\langle Q \rangle &=&\frac{\left\langle  P \right\rangle}{M }dt+ 2  du_S \sigma_{QQ}
\\
\\
d\left\langle P \right\rangle &=& - M\omega^2_0 \langle Q \rangle  dt 
- dt \left\langle  B \right\rangle 
+ 2 du_S \sigma_{PQ} 
-\hbar dv_S 
\\ 
\\
d\sigma_{QQ} &=& 2\frac{dt}{M} \sigma_{PQ} 
 \\
 \\
d\sigma_{PP} &=& 
-2 M\omega^2_0 dt \sigma_{PQ} \\
\\
d \sigma_{PQ} &=&  \frac{dt}{M} \sigma_{PP} -M\omega_0^2 \sigma_{QQ} dt  
\end{array}
\right.
\label{eq:mom}
\end{eqnarray}
Above stochastic equations differ from the standard Langevin Eqs. used to treat dissipation. Indeed, (i) the stochastic noise 
appear in both momentum and position evolutions. (ii) since the noise is complex, the expectation values of 
$\langle Q \rangle$ and $\langle P \rangle$ are also complex (see left side of Fig. \ref{fig:fusion1}). 
This points out that the densities are not Hermitian along the 
stochastic paths, a property which also differs from more standard treatment of open quantum systems. 
Concentrating now on second moments, quantum fluctuations, denoted by $\sigma$ here, are automatically accounted for in the present theory. Total  
fluctuation of a given observable, denoted by $\Sigma$ are obtained by summing up statistical fluctuations on top of quantum fluctuations along each paths. The evolution of different fluctuations are illustrated in right side of Fig. \ref{fig:fusion1}, the quantum fluctuations (dashed line) differs significantly from the exact solution (solid line), while 
the total quantum+statistical fluctuations (filled circles) are in very good agreement with the exact dynamics.
The different contributions to the fluctuations of $Q$ are illustrated in right side of Fig. 
\ref{fig:fusion1}.    
\begin{figure}[t!]
\includegraphics[height=5.cm]{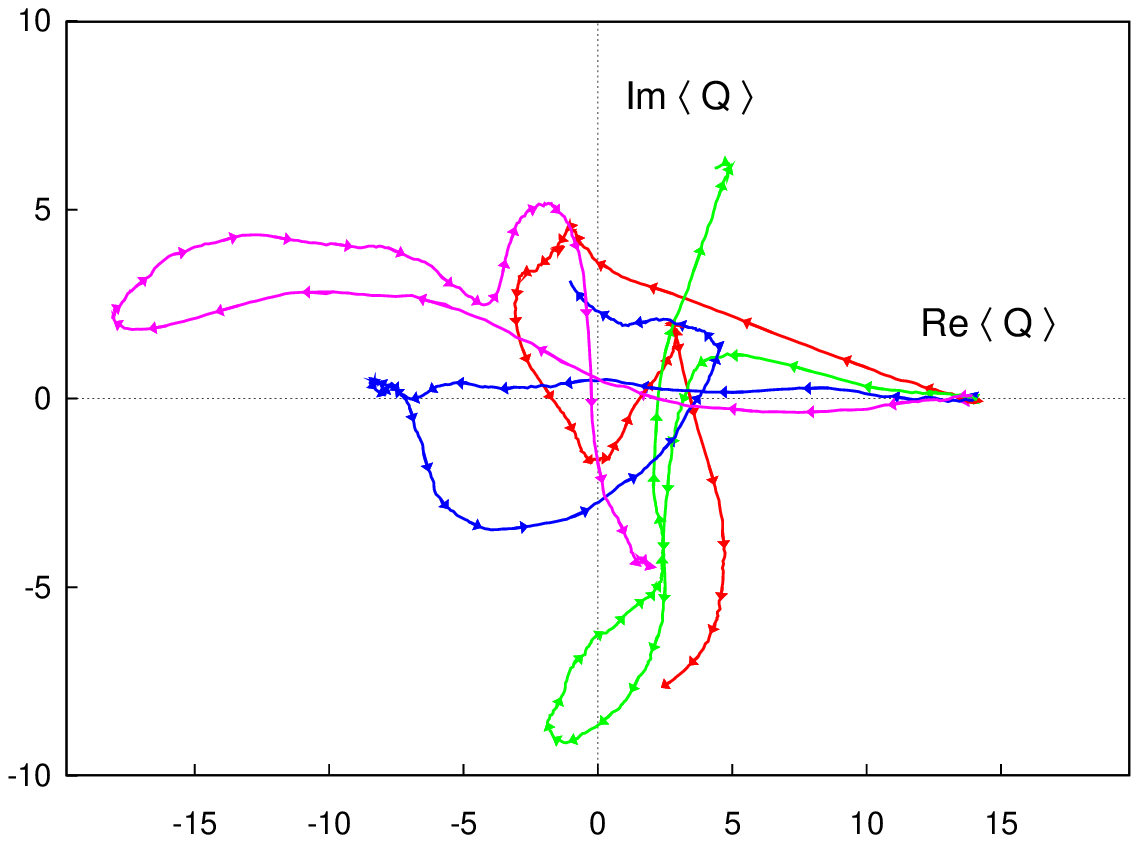}
\includegraphics[height=5.cm]{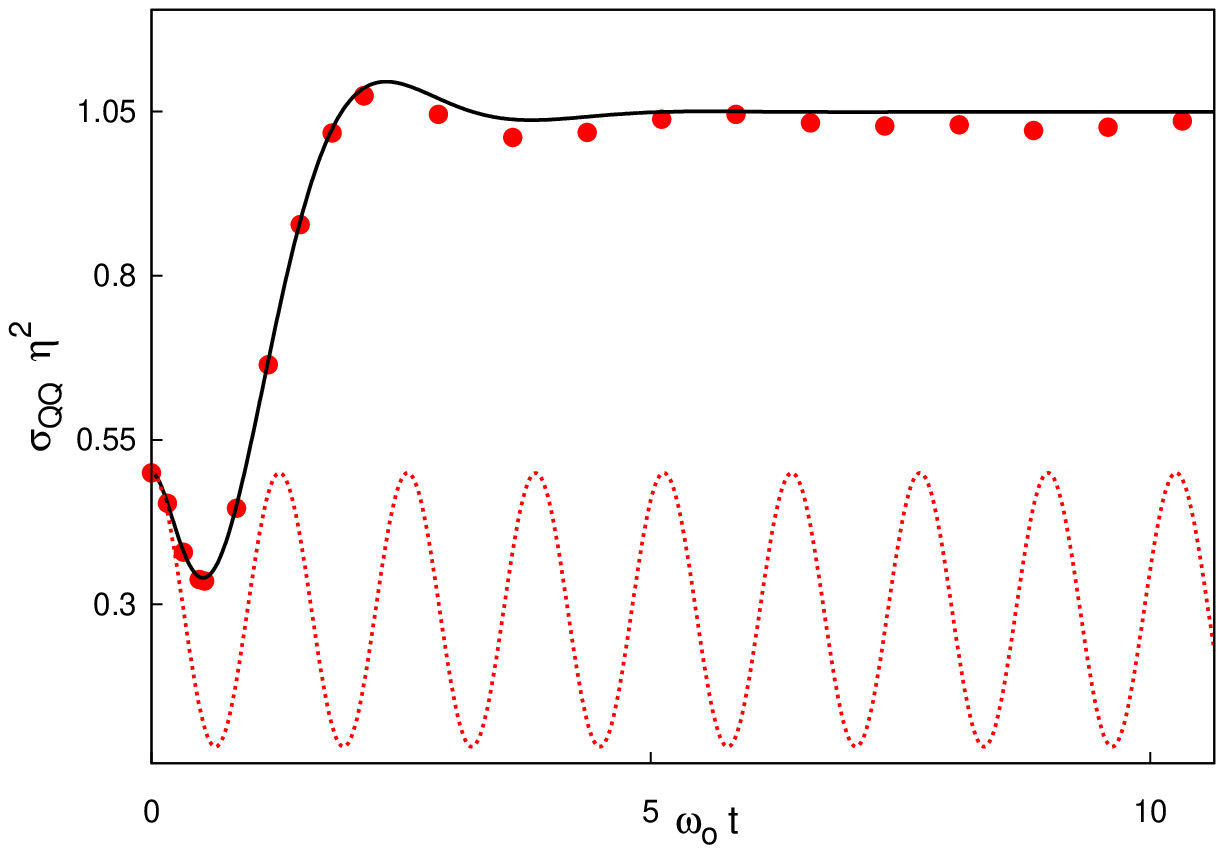}
\caption{Left: Three examples of $\langle Q \rangle$ stochastic evolutions in the complex plane. Right: Exact evolution of $\Sigma_{QQ}$ 
(solid line) compared to the quantum (only) second moment evolution (dashed line) and quantum+statistical (filled circles).
Results are obtained with $7 \times 10^5$ trajectories assuming a spectral function of the form $\displaystyle J(\omega) = m \eta 
\omega \Delta^2_c /(\Delta^2_c + \omega^2)$ 
with  $\Delta_c = 5 \omega_0$, $\eta = 0.5 \hbar\omega_0$ while $ k_B T = \hbar\omega_0$.} 
\label{fig:fusion1}
\end{figure}

\section{Summary and discussion on the Many-Body problem}

In this proceeding, the possibility to treat exactly the evolution of a system coupled to a complex environment has been 
illustrated in the Caldeira-Leggett model. We have shown, that the present technique can provide a very powerful tool to treat 
dissipative processes including non-Markovian effects. More illustration will be found 
in ref. \cite{Hup08}. 

There is a close analogy between open systems and closed Many-Body systems
treated approximately by selecting few relevant degrees of freedom. Then, the irrelevant 
degrees of freedom act as an environment
for the observables under interest \cite{Abe96,Lac04}. 
For instance, numerous aspects of strongly interacting systems have been understood by replacing 
the initial complex problem by an independent particle problem. This amount to focus essentially  on one-body 
degrees of freedom which is conveniently made by introducing mean-field theories. It has been shown 
recently that stochastic mean-field theories can be introduced to not only describe one-body quantities but 
also two-, three- or higher order observables. Simple applications based on the concept of
Stochastic Schr\"odinger Equation (SSE) have revealed difficulties due to unstable trajectories, a problem often encountered
when solving non linear stochastic equations \cite{Gar00}. We have observed that 
similar difficulties occur when using SSE in open quantum systems. However, the direct use of observables evolution (Eqs. (\ref{eq:mom})) seems to cure this problem. Although the numerical implementation of observables evolution might be more demanding than solving 
Schr\"odinger equations, the success of the former in open quantum systems 
is rather encouraging to overcome recent difficulties observed in many-body problems.      




\bibliographystyle{aipproc}   


\begin{thebibliography}{99}


\bibitem{Lac08}{D. Lacroix, Phys. Rev. {\bf E77}, 041126 (2008).}

\bibitem{Abe96} {Y. Abe, S. Ayik, P.-G. Reinhard and E. Suraud, Phys. Rep. {\bf 275}, 49 (1996).}

\bibitem{Lac04}  D. Lacroix, S. Ayik and Ph. Chomaz, Progress in Part. and
Nucl. Phys. {\bf 52} (2004) 497.

\bibitem{Lac07} {D. Lacroix, Ann. of Phys. {\bf 322}, 2055 (2007).}

\bibitem{Sha04} 
J. Shao, J. Chem. Phys. {\bf 120}, 5053 (2004).
Y. Yan,F. Yang, Y. Liu and J. Shao, Chem. Phys. Lett. {\bf 395}, 216 (2004).

\bibitem{Lac05} D. Lacroix, Phys. Rev. {\bf A72}, 013805 (2005). 

\bibitem{Cal83} A. O. Caldeira, A. J. Leggett, Ann. of Phys. {\bf 149}, 374 (1983).

\bibitem{Bre02} H.P. Breuer and F. Petruccione, \textit{The Theory of Open
Quantum Systems} (Oxford University Press, Oxford, 2002).



\bibitem{Hup08}{G. Hupin and D. Lacroix, {\it in preparation}.} 

\bibitem{Gar00} W. Gardiner and P. Zoller, ''\textit{Quantum Noise}'',
Springer-Verlag, Berlin-Heidelberg, 2$^{nd}$ Edition (2000).



\end{thebibliography}

\IfFileExists{\jobname.bbl}{}
 {\typeout{}
  \typeout{******************************************}
  \typeout{** Please run "bibtex \jobname" to optain}
  \typeout{** the bibliography and then re-run LaTeX}
  \typeout{** twice to fix the references!}
  \typeout{******************************************}
  \typeout{}
 }


\end{document}